\documentclass[%
 aip,
jap,%
 amsmath,amssymb,
 reprint,%
footinbib,
citeautoscript]{revtex4-1}

\usepackage{graphicx}
\usepackage{dcolumn}
\usepackage{bm}
\usepackage{tabularx}
\usepackage{eso-pic}
\usepackage{fix-cm}
\usepackage{natbib}
\usepackage{subfigure}

\newcommand{\sss}{\scriptscriptstyle}
\newcommand{\sst}{\scriptstyle}

\newcommand{\stext}[1]{\sss \text{#1} \sst}

\makeatother

\begin{document}

\title{Infrared dielectric functions, phonon modes and free-charge carrier properties of high-Al-content Al$_{x}$Ga$_{1-x}$N alloys determined by mid-infrared spectroscopic ellipsometry and optical Hall effect}

\author{S.~Sch\"{o}che}
\affiliation{Department of Electrical, and Computer Engineering, and Center for Nanohybrid Functional Materials, University of Nebraska-Lincoln, Lincoln, 68588-0511, U.S.A.} 
\author{T.~Hofmann} \affiliation{Department of Electrical, and Computer Engineering, and Center for Nanohybrid Functional Materials, University of Nebraska-Lincoln, Lincoln, 68588-0511, U.S.A.}  \affiliation{{Department of Physics, Chemistry and Biology (IFM), Link\"oping University, SE-58183, Sweden}}
\author{D.~Nilsson} \affiliation{{Department of Physics, Chemistry and Biology (IFM), Link\"oping University, SE-58183, Sweden}}
\author{A.~Kakanakova-Georgieva} \affiliation{{Department of Physics, Chemistry and Biology (IFM), Link\"oping University, SE-58183, Sweden}}
\author{E.~Janz\'en} \affiliation{{Department of Physics, Chemistry and Biology (IFM), Link\"oping University, SE-58183, Sweden}}
\author{P.~K\"{u}hne} \affiliation{{Department of Physics, Chemistry and Biology (IFM), Link\"oping University, SE-58183, Sweden}}
\author{K.~Lorenz} \affiliation{{IPFN, Instituto Superior T\'ecnico, Universidade de Lisboa, Estrada Nacional 10, 2695-066 Bobadela LRS, Portugal}}
\author{M.~Schubert} \affiliation{Department of Electrical, and Computer Engineering, and Center for Nanohybrid Functional Materials, University of Nebraska-Lincoln, Lincoln, 68588-0511, U.S.A.} 
\author{V.~Darakchieva} 
\email{vanya.darakchieva@liu.se} \affiliation{{Department of Physics, Chemistry and Biology (IFM), Link\"oping University, SE-58183, Sweden}}

\begin{abstract}

The phonon mode parameters and anisotropic mid-infrared dielectric function tensor components of high-Al-content Al$_{x}$Ga$_{1-x}$N alloys in dependence of the Al content $x$ are precisely determined from mid-infrared spectroscopic ellipsometry measurements for a set of high-quality Si-doped Al$_{x}$Ga$_{1-x}$N epitaxial layers on 4H-SiC substrates. Two-mode behavior of the $E_1$(TO) modes and one-mode behavior of the $A_1$(LO) mode are found in agreement with previous Raman scattering spectroscopy reports. The composition dependencies of the IR active phonon frequency parameters are established and a discussion on the silent $B_1$ mode that may be disorder activated is provided. The static dielectric constants in dependence of $x$ are determined by using the best-match model derived phonon mode frequency and high-frequency dielectric constant parameters and applying the Lydanne-Sachs-Teller relation. The effective mass parameter in high-Al-content Al$_{x}$Ga$_{1-x}$N alloys and its composition dependence are determined from  mid-infrared optical Hall effect measurements. Furthermore, the free electron concentration $N$ and mobility  parameters $\mu$ of Al$_{x}$Ga$_{1-x}$N films with similar Si doping levels are investigated as function of the Al content, $x$ and discussed. 
\end{abstract}

\date{\today}


\maketitle

\section{Introduction}\label{Sec:Intro}

Al$_{x}$Ga$_{1-x}$N alloys with high molar fraction of Al (high-Al-content Al$_{x}$Ga$_{1-x}$N) are highly interesting as active layers in optoelectronic devices operated in the deep-ultraviolet (DUV) spectral range, including UV light emitting diodes and high-power UV laser diodes. Achieving reliable n-type conductivity in high-Al-content AlGaN and AlN is very challenging due to the high ionization energy of the common donors (O and Si)~\cite{Neuschl2013}. Successful n-type doping of high-Al-content AlGaN and insights on the shallow/DX behavior of the Si donor has recently been reported~\cite{Neuschl2013,Son2011,Kakanakova-Georgieva2013,Trinh2014}. Understanding the device properties and optimization of device design requires detailed knowledge of the basic electronic properties of the material, including the effective mass parameter. The electrical characterization of these properties is often impeded by difficulties to form ohmic contacts. Optical methods like photoluminescence or cathodoluminescence rely on excitation sources of higher energy than the fundamental energy band gap of the investigated material and allow the estimation of the reduced free-charge carrier (FCC) effective mass parameter only if additional input such as the hole effective mass parameter is assumed. IR reflectometry and Raman scattering measurements allow estimates of the electron effective mass parameter from experimentally determined plasma frequencies, but also rely on precise electrical data, e.g., from electrical Hall effect measurements, and are not sensitive to individual layers of doped material within multilayer heterostructures. Note, that complex strain-engineered multilayer structures are required in order to fabricate low-defect density crack-free AlGaN layers, in particular with high Al content.

Mid-infrared (MIR) spectroscopic ellipsometry (MIR-SE) allows access to the dielectric function (DF) tensor of optically anisotropic materials. Relevant physical parameters such as phonon mode frequency and broadening parameters can be obtained from a lineshape analysis of the determined MIR DF by using parameterized model DFs (MDF). The FCC parameters concentration $N$ and mobility $\mu$ can be extracted from the plasma frequency and broadening parameters by applying the classical Drude model if the effective mass parameter is known~\cite{SchubertIRSEBook_2004,Kasic2000,DarakchievaICSE4}. The optical Hall effect (OHE), which comprises generalized spectroscopic ellipsometry measurements in combination with external magnetic fields, provides access to the energy-distribution-averaged FCC parameters concentration, mobility, and effective mass~\cite{Schubert2003a, Hofmann2003,Schubert2004a, Hofmann2006a, Hofmann2006, Hofmann2008a, DarakchievaAPL09, Schoeche2011}. Previous investigations on GaAs~\cite{Schubert2003a,Hofmann2007}, AlGaInP and BInGaAs alloys~\cite{Schubert2004a}, AlInP~\cite{Hofmann2008a}, ZnMnSe~\cite{Hofmann2006,Hofmann2006b}, InN~\cite{Hofmann2006,Hofmann2008,DarakchievaAPL09}, AlGaN/GaN high-electron mobility transistor structures~\cite{Schoeche2011,Hofmann2012}, and graphene~\cite{Hofmann2011} showed that the OHE provides accurate values of the effective mass parameter that corroborate the results found, for example, from Shubnikov-de Haas measurements. Accurate knowledge of the MIR DF is a prerequisite for the analysis of the OHE data, for investigation of FCC properties in doped, high-Al-content Al$_{x}$Ga$_{1-x}$N alloys by MIR-SE/reflectometry, and for MIR-SE/reflectometry studies of complete device structures.

For GaN, AlN, and AlGaN alloys with wurtzite crystal structure, an anisotropy of the electron effective mass parameter parallel and perpendicular to the $c$-axis is expected~\cite{Rinke2008}. From MIR-SE in combination with electrical Hall effect measurements, slightly anisotropic electron effective mass parameters of $m^\ast_\parallel=(0.228\pm0.008)\,m_0$ parallel to the $c$-axis and $m^\ast_\perp=(0.237\pm0.006)\,m_0$ perpendicular to the $c$-axis were previously reported for GaN~\cite{Kasic2000,Feneberg2013}. For increasing Al-content in AlGaN alloys, an increasing value of the effective mass parameters is expected. Xu~\emph{et al.} demonstrated an increase of the transverse effective mass parameter (perpendicular to $c$-axis) from $m^\ast=0.27\,m_0$ to $m^\ast=0.30\,m_0$ with increasing Al-content for $x$ between 0 and 0.52 by combining electrical Hall effect measurements with IR reflectivity measurements~\cite{Xu2005}. Recently, we reported an isotropically averaged effective mass parameter of $m^\ast=(0.336\pm0.020)\,m_0$ for Al$_{0.79}$Ga$_{0.21}$N with no significant anisotropy determined from the MIR-OHE measurements.\cite{SchocheAPL13} \footnote{Note, that the composition value $x=\,0.72$ for the sample investigated in Ref.~\citenum{SchocheAPL13} refers to the value as determined by SIMS. The apparent Al content as determined from XRD in this sample is $x=\,0.79$. The apparent content used in the current work was determined by complementary XRD and RBS measurements, which were found to be in excellent agreement. As already reported, the Al content extracted from XRD is equivalent to the SIMS determined Al content within 0.05.\cite{Nilsson2014} Differences between the apparent XRD content and the one determined by SIMS are observed which may be attributed to the use of standards in SIMS, complicated strain relaxation in the films, uncertainties in the strain-free lattice parameters of AlN and GaN, deviations from Vegard's rule in the lattice and stiffness constants, etc.  Note, that the results reported in Ref. ~\citenum{SchocheAPL13} are still valid, but the extrapolated values of the effective mass of AlN needs to be rescaled according to the adjusted apparent content determined by XRD\label{footnote1}}


The lattice dynamics of Al$_x$Ga$_{1-x}$N alloys over the entire composition range were reported by Davydov~\emph{et al.}~\cite{Davydov2002} First- and second-order Raman scattering in hexagonal Al$_x$Ga$_{1-x}$N alloy epilayers grown by molecular beam epitaxy on top of thin GaN buffer layers on $c$-plane sapphire substrates (0.5\,$\mu$m thick, Al-content $0<x<0.5$) and by reduced-pressure metal-organic chemical vapor deposition on thick ($1-4\,\mu$m) Al$_x$Ga$_{1-x}$N nucleation layers deposited at low-temperature on sapphire substrates ($2-4\,\mu$m thick, Al-content $0.5<x<1$) were investigated. One-mode behavior of the $A_1$(LO) and $E_1$(LO) phonons and two-mode behavior of the $E_2$(high), $E_1$(TO), $E_2$(low), and $A_1$(TO) phonons was reported for the entire composition range. According to the energetic position compared to the pure binary alloys GaN and AlN, the low-frequency $E_1$(TO) mode is referred to as the GaN-like mode while the high-frequency $E_1$(TO) mode is referred to as AlN-like mode. The energetic position of the $B_1$(high) silent mode has been proposed from an analysis of the first- and second-order Raman spectra of the Al$_{x}$Ga$_{1-x}$N alloys. The results presented by Davydov~\emph{et al.} were confirmed for subsets of the composition range by means of MIR-SE~\cite{Schubert1999,Kasic03} ($x=0-0.5$), by IR reflectometry~\cite{Yu1998,Wisniewski1998}, and Raman scattering~\cite{Holtz2001,Kim2011,Kirste2012} and agree with theoretical predictions~\cite{Grille2000}. However, the samples investigated by Davydov~\emph{et al.} were grown by different growth techniques and without strain-engineering. No information on the structural properties of these films were reported in Ref.~\citenum{Davydov2002}, however, the high-Al-content films were likely cracked. 
The set of crack-free samples investigated in this work presents an excellent case study of the compositional dependencies of AlGaN phonon and FCC parameters. Wavelength-by-wavelength extracted data or parameterized model dielectric function (MDF) tensor components in the MIR spectral range for high-Al-content AlGaN alloys have not been reported so far. Further, the static dielectric constant for high Al-contents which is needed for various calculations has not been experimentally accessed. 

In this work, the MIR DF, phonon and FCC parameters of high-Al content Al$_{x}$Ga$_{1-x}$N epitaxial layers are determined as a function of Al content. The compositional dependence of the effective mass parameter of Si-doped Al$_{x}$Ga$_{1-x}$N is established and used to estimate the effective mass parameter of AlN. Further, the effect of Al content on the FCC concentration and mobility parameters in Al$_{x}$Ga$_{1-x}$N with similar doping Si concentrations is investigated.

\section{Experimental}

A set of Si-doped, crack-free and low defect density epitaxial wurtzite Al$_{x}$Ga$_{1-x}$N films with nominal thicknesses between 400\,nm and 1200\,nm was grown by metal-organic chemical vapor deposition (MOCVD)  on on-axis semi-insulating 4H-SiC substrates. The Si doped layers were grown on effective template consisting of an undoped AlN layer, a composition graded AlGaN layer, and an undoped Al$_{x}$Ga$_{1-x}$N layer. Details on growth conditions and structural properties of the samples can be found in Refs.~\onlinecite{Nilsson2014} and \onlinecite{Nilsson2015}, whereby  secondary ion mass spectrometry (SIMS) was used to determine Al content and Si concentration; the net donor concentration ($N_\mathrm{D}-N_\mathrm{A}$) was estimated by capacitance-voltage (C-V) measurements. The Al content in the films was also determined by high-resolution x-ray diffraction (HR-XRD) following Ref. \onlinecite{DarakchievaJAP08} and Rutherford back scattering spectrometry (RBS), and referred hereafter as the apparent Al content. The apparent Al content in the Al$_{x}$Ga$_{1-x}$N layers was varied from 0.61 to 1. An exemplary HR-XRD reciprocal space map around the asymmetric AlGaN (10$\bar{1}$5) reciprocal lattice point is shown in Fig.~\ref{fig:XRD} left panel. High crystalline quality is manifested by the small broadening of the individual Al(Ga)N reciprocal lattice points. 

MIR-SE measurements have been carried out at room temperature using a commercial Fourier transform-based MIR ellipsometer (J.A.~Woollam Co.,~Inc.) in the spectral range from 300\,cm$^{-1}$ to 6000\,cm$^{-1}$ with a resolution of 2\,cm$^{-1}$, and at angles of incidence $\varPhi_{\stext{a}}=50,70^{\circ}$. A custom-built Fourier transform-based, rotating-analyzer MIR ellipsometer was used for the MIR-OHE measurements in the spectral range from 600\,cm$^{-1}$ to 1700\,cm$^{-1}$ with a resolution of 1\,cm$^{-1}$~\cite{Kuehne2014}. The MIR-OHE measurements were carried out at room temperature at an angle of incidence $\varPhi_{\stext{a}}=45^{\circ}$. The measurements were performed at $-7$\,T, $0$\,T and $+7$\,T, with the magnetic field oriented parallel to the incoming beam, resulting in a magnetic field strength $B_c=B/\sqrt{2}$ along the sample normal~\cite{Kuehne2014}.

\section{Theory}

\subsection{Optical phonons in Al$_{x}$Ga$_{1-x}$N}

GaN, AlN, and Al$_{x}$Ga$_{1-x}$N alloys crystallize in the wurtzite structure with space group C$^6_{6v}$. For the binary compounds, group theory predicts the irreducible representation $\Gamma_{\mathrm{ac}}+\Gamma_{\mathrm{opt}}=(\mathrm{A}_1+\mathrm{E}_1)+(\mathrm{A}_1+2\mathrm{B}_1+\mathrm{E}_1+2\mathrm{E}_2)$. Only the optical phonons of A$_1$ and E$_1$ symmetry are accompanied by a dipole moment with electric field vector $\boldsymbol{E}\perp c$ (E$_1$ mode) and $\boldsymbol{E}\parallel{c}$ (A$_1$ mode), respectively. These modes are both Raman and IR active, while the E$_2$ modes are only Raman active, and the B$_1$ modes are silent. The phonon modes of the binary compounds GaN and AlN were studied extensively by Raman scattering and IR reflection/ellipsometry and are well established. 

\begin{figure} 
	\centering
			\includegraphics[keepaspectratio=true,trim= 20 0 0 0, clip, width=0.3\textwidth]{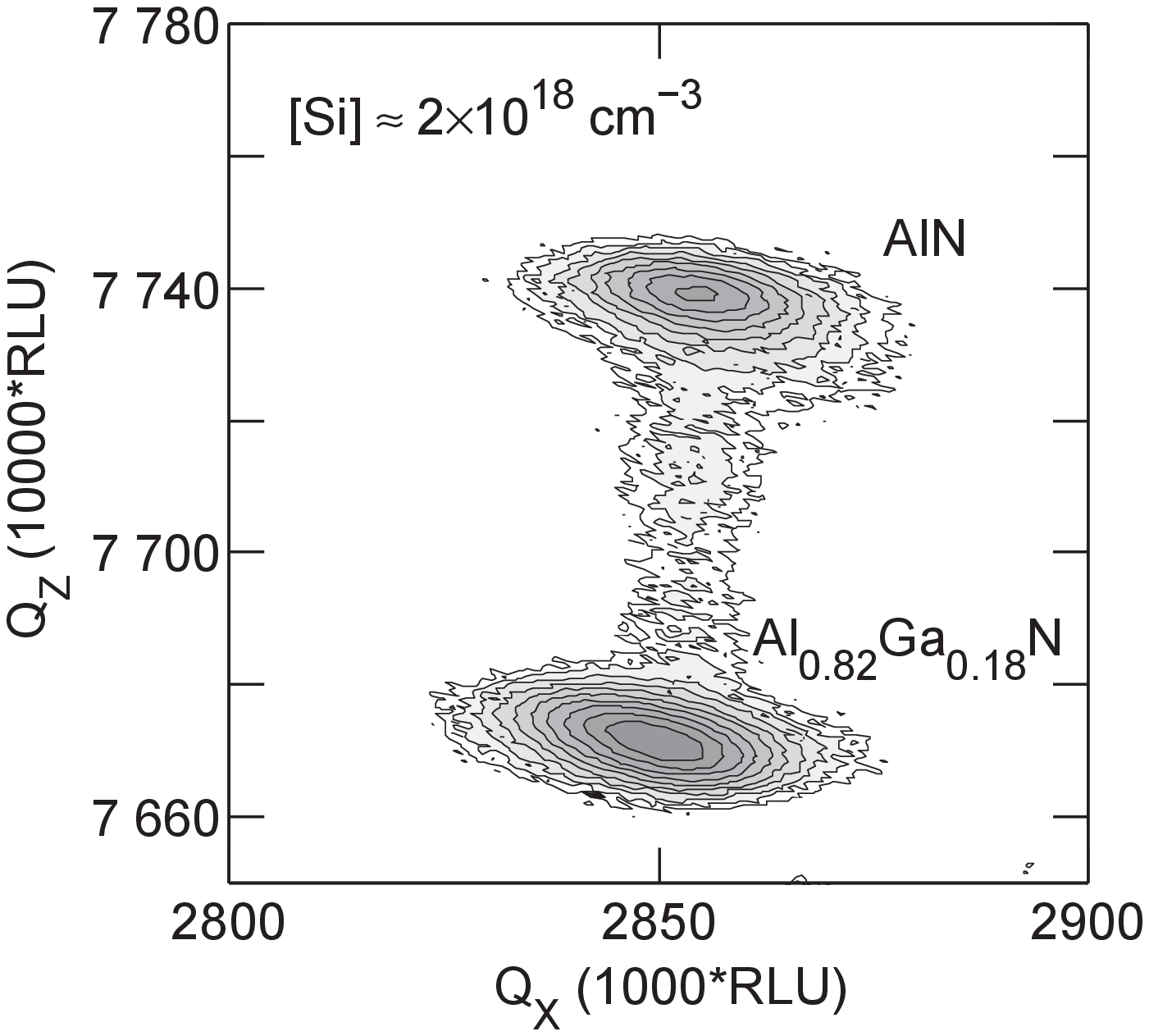} 
			\includegraphics[keepaspectratio=true,trim= 0 -200 0 0, clip, width=0.17\textwidth]{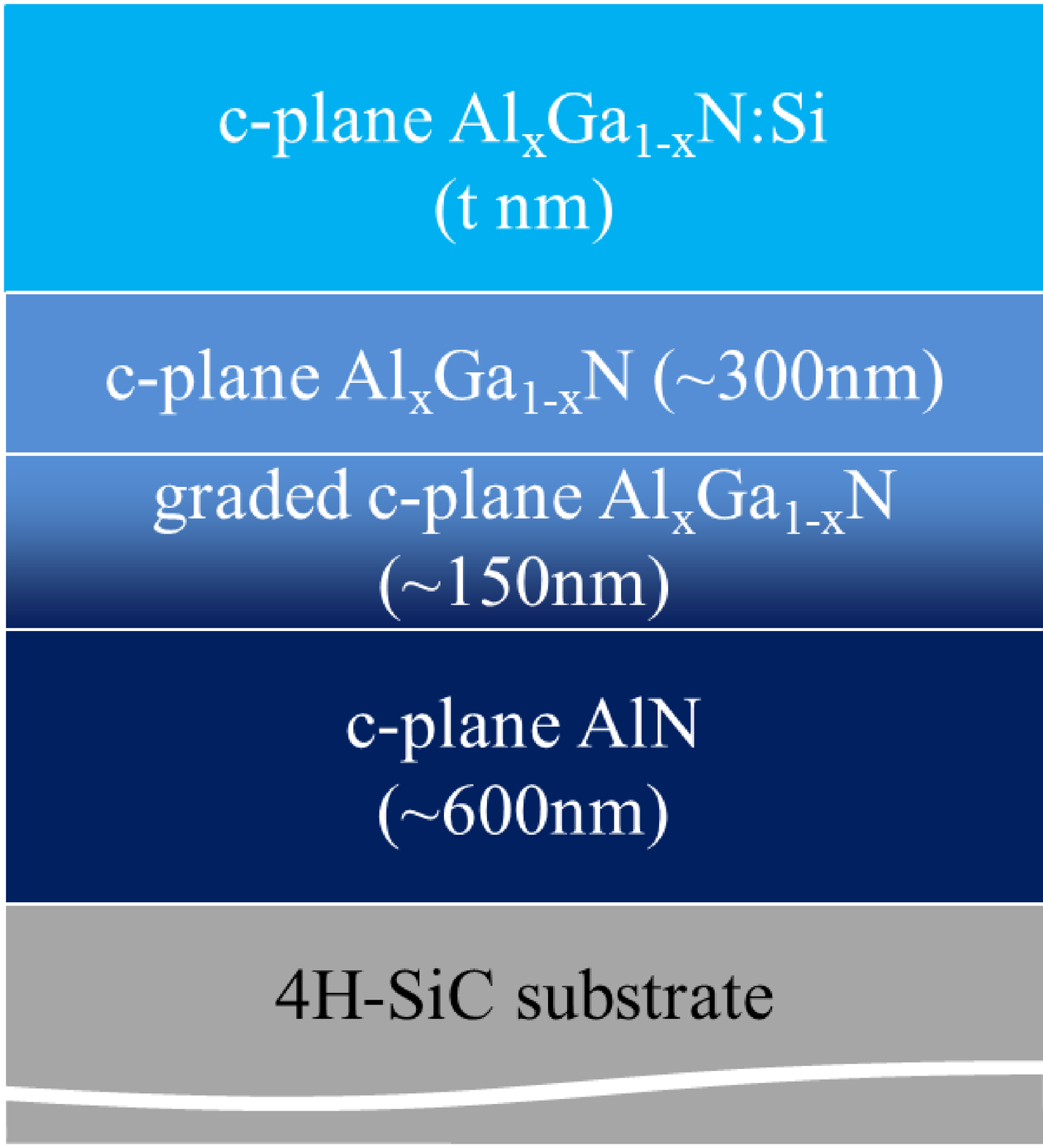} 
			\caption{left panel: Exemplary XRD reciprocal space map around the asymmetric AlGaN (10$\bar{1}$5) reciprocal lattice point for a Si-doped Al$_{0.82}$Ga$_{0.18}$N sample. right panel: Schematic drawing of the layer stack model used for the model analysis.} 
	\label{fig:XRD}
\end{figure}

Optical phonons in randomly mixed ternary alloys A$_{1-x}$B$_x$C can be categorized into two main classes referred to as one-mode and two-mode behavior. In the case of one-mode behavior, only one set of TO and LO frequencies is found which change continuously and nearly linearly with $x$ from the AC component to the BC component and show an approximately constant strength of each mode. The term two-mode behavior is used for situations in which two energetically well separated sets of phonons are observed for a given composition $x$, which occur at frequencies close to those of the binary compounds AC and BC. The oscillation strength of each set of phonons is approximately proportional to the content of the component it is associated with. For the Al$_{x}$Ga$_{1-x}$N alloys, two-mode behavior over the whole composition range for the E$_1$(TO) is consistently reported in IR reflection/ellipsometry and Raman scattering studies~\cite{Davydov2002,Schubert1999,Kasic03,Yu1998,Wisniewski1998,Holtz2001,Kim2011,Kirste2012}. For the A$_1$(TO), A$_1$(LO), and E$_1$(LO) phonon modes, one-mode behavior was reported from Raman scattering studies~\cite{Davydov2002,Holtz2001,Kim2011,Kirste2012}.

\subsection{Ellipsometry model and data analysis}

Spectroscopic ellipsometry is an indirect method and a detailed model analysis is required in order to extract relevant physical parameters. A stratified layer model analysis is employed to analyze the SE data set. All sample constituents discussed here have hexagonal or wurtzite crystal structures and are therefore optically anisotropic with uniaxial optical properties. In all samples, the optical axes of all constituents are oriented along the sample surface normal. Thus, without applying a magnetic field during the measurement, no mode conversion of light polarized parallel (p) to polarization perpendicular (s) to the plane of incidence and vice versa occurs and the standard ellipsometry formalism can be applied. The standard ellipsometric parameters $\Psi$ and $\Delta$ are defined by the ratio $\rho$ of the complex valued Fresnel reflection coefficients
\begin{equation}
	\rho=\frac{r_\mathrm{p}}{r_\mathrm{s}}=\tan{\Psi}\exp(\mathrm{i}\Delta).
\end{equation}

The OHE comprises spectroscopic ellipsometry measurements in combination with external magnetic fields. In the presence of FCC, magnetic-field-induced birefringence occurs and the generalized ellipsometry formalism needs to be applied. Generalized ellipsometry determines the dielectric tensor of each sample constituent. The interaction between the electromagnetic plane wave and the sample is conveniently represented using the Mueller matrix calculus. In this formalism, the optical response of the birefringent sample is characterized by the $4\times4$ Mueller matrix which relates the real-valued Stokes vectors before, $S^{\mathrm{in}}$, and after interaction with a sample~\cite{Schubert04}: $S^{\mathrm{out}}=MS^{\mathrm{in}}$.

The ellipsometry data was analyzed by using a layer-stack model 4H-SiC/AlN buffer/graded AlGaN/undoped Al$_{x}$Ga$_{1-x}$N/ Si-doped Al$_{x}$Ga$_{1-x}$N  (see Fig.~\ref{fig:XRD}, right panel). The light propagation within the entire sample stack is calculated by applying a $4\times4$ transfer matrix algorithm for multilayer systems assuming plane parallel interfaces~\cite{SchubertIRSEBook_2004}. The DF tensor components of each sample constituent are needed as input parameters to perform the layer model calculations. These DF tensor components can be given as tabulated values for each wavelength or can be modeled by using parameterized, wavelength-dependent algebraic MDF. The MDF describe a specific physical process in the material such as a lattice vibration or absorption by FCC and are combined in order to render the overall spectral shape of a material's DF. 

The MDFs of the 4H-SiC substrate and AlN were determined from
MIR-SE measurements on a bare substrate and AlN/SiC, respectively, and used without any changes during the model analysis. The AlGaN MDFs were parameterized using contributions from optically active phonon modes $\varepsilon^{\text{L}}(\omega)$ and FCC excitations $\varepsilon^{\text{FC}}(\omega)$. In the model, additional FCC contributions were allowed in the SiC substrate and the AlN buffer layer, but no FCC contributions from these layers were revealed during the regression analysis. The functions $\varepsilon^{\text{L}}_j(\omega)$ are parameterized with Lorentzian lineshapes which account for transverse (TO) and longitudinal optic (LO) phonon frequencies, $\omega_{\stext{TO,}j}$ and $\omega_{\stext{LO,}j}$, respectively, for polarization $j$ =``$\parallel$'', ``$\perp$'' to the crystal $c$-axis \cite{SchubertIRSEBook_2004}:
\begin{equation}
\varepsilon^\mathrm{L}_j(\omega)=\varepsilon_{\infty,j}\prod\limits_{l}^{k}\frac{\omega^2+\mathrm{i}\gamma_{\mathrm{LO},lj}\omega-\omega^2_{\mathrm{LO},lj}}{\omega^2+\mathrm{i}\gamma_{\mathrm{TO},lj}\omega-\omega^2_{\mathrm{TO},lj}}.
\label{Eq:IR-model}
\end{equation}
$\omega^2_{\mathrm{LO},l\parallel}$, $\omega^2_{\mathrm{TO},l\parallel}$, $\omega^2_{\mathrm{LO},l\perp}$, and $\omega^2_{\mathrm{TO},l\perp}$ indicate the frequencies of the A$_1$(LO)$_l$, A$_1$(TO)$_l$, E$_1$(LO)$_l$, and the E$_1$(TO)$_l$ phonon modes of the wurtzite group-III nitride compounds, respectively. The thin graded AlGaN layer was accounted for in the model analysis by introducing a linearly graded layer (10 slices) for which the phonon mode parameters at the bottom and top of the layer were fixed to the parameters of the adjacent AlN and Al$_{x}$Ga$_{1-x}$N, respectively. Identical phonon mode parameters were used for the undoped and Si-doped Al$_{x}$Ga$_{1-x}$N layer.

In the spectral range between TO and LO phonon modes, the material becomes totally reflecting (reststrahlen range). Distinct features in the spectra close to these phonon modes allow precise determination of the phonon mode frequency and broadening parameters. $\varepsilon_{\infty,j}$ is the high-frequency dielectric constant which accounts for offsets in the real part of the DF caused by electronic interband transition at higher energies outside the measured spectral range. In this highly symmetric form, the DF $\varepsilon^{\text{L}}_j(\omega)$ and dielectric loss function $\varepsilon^{\text{L}}_j(\omega)^{-1}$ can easily be evaluated. The energies $\omega_{\mathrm{TO}{i}}$ are the poles of the DF, while the energies $\omega_{\mathrm{LO}{i}}$ are given by the poles of the dielectric loss function. The static dielectric constant $\varepsilon_{0,j}=\varepsilon^\text{L}_j(\omega=0)$ is related to the high-frequency dielectric constant by the Lyddanne-Sachs-Teller (LST) relation~\cite{Lyddane1941,Cochran1962}
\begin{equation}
\varepsilon_{0,j}=\varepsilon_{\infty,j}\prod\limits_{l}^{k}\frac{\omega_{\mathrm{LO},lj}^2}{\omega_{\mathrm{TO},lj}^2}.
\label{Eq:LST}
\end{equation}
The factorized phonon mode model in Eq.~\ref{Eq:IR-model} is a generalization of the classical damped harmonic oscillator and allows for different broadening parameters of the TO and LO modes.  Note that different phonon decay times should be considered for TO and LO phonons for multiple phonon-mode crystals of large TO-LO splitting~\cite{Gervais1973,Gervais1974a}.

$\varepsilon^{\text{FC}}_j(\omega)$ is parameterized using the classical Drude formalism \cite{Pidgeon80}
\begin{equation}
\varepsilon^{\text{FC}}_j(\omega)=-\varepsilon_{\infty,j}\frac{\omega^2_{\mathrm{p},j}}{\omega(\omega+\mathrm{i}\gamma_{\mathrm{p},j})},
\label{Eq:Drude}
\end{equation}
The screened plasma frequency $\omega_{\mathrm{p},j}$ and the plasma broadening parameter $\gamma_{\mathrm{p},j}$ are related to the FCC concentration $N$, the effective mass parameter $m_j^\ast$, and the optical mobility parameter $\mu_j$ by 
\begin{equation}
\omega^2_{\mathrm{p},j}=-\frac{Ne^2}{m^\ast_j\varepsilon_0\varepsilon_{\infty,j}} \ \mathrm{and} \ \gamma_{\mathrm{p},j}=\frac{e}{m^\ast_j\mu_j},
\label{Eq:plasma}
\end{equation}
under assumption of energy-independent scattering mechanisms. $\varepsilon_0$ is the vacuum permittivity, $\varepsilon_{\infty,j}$ denotes the high frequency dielectric constant and $e$ is the elementary charge. The plasma frequency $\omega_{\mathrm{p},j}$ and plasma broadening parameter $\gamma_{\mathrm{p},j}$ provide access only to the coupled quantities $N/m^{\ast}_j$ and $\mu_j{m^{\ast}_j}$. Conclusions about carrier concentration and mobility from the MIR-SE measurements are therefore only possible by assuming a value for the effective mass of electrons or holes, respectively.

The DF tensor for FCC in external magnetic field (extended Drude model) is derived from the equation of motion and is given by an asymmetric tensor which allows for determination of the screened plasma frequency tensor $\bm{\omega}_{\stext{p}}$ and the cyclotron frequency tensor $\bm{\omega}_{\stext{c}}$\cite{Pidgeon80,Hofmann2008a}:
\begin{align}\label{DA}
&\bm{\varepsilon}^{\text{FC}}(\omega)  = \bm{\omega}_{\mbox{{\tiny
p}}}^{2}\\\nonumber
&\times\left[ -\omega^{2}\bm{I}-i\omega\bm{\gamma}+i\omega
\left(\begin{array}{ccc}
0 & b_{3} & -b_{2}\\
-b_{3} & 0 & b_{1}\\
b_{2} & -b_{1} & 0
\end{array}\right)
\bm{\omega}_{\mbox{\tiny c}} \right]^{-1}.
\end{align}
The inverse FCC mobility tensor is given by $(\bm{\mu})^{-1}= \bm{\gamma}\bm{m}^{\ast}/q$ where $\bm{m}^{\ast}$ denotes the effective mass tensor in units of the free electron mass $m_{0}$.  The plasma frequency tensor $\bm{\omega}_{\stext{p}}$ is related to the FCC density parameter $N$ and the effective mass tensor $\bm{m}^{\ast}$ by $\bm{\omega}_{\stext{p}}^2=Nq^2/(\varepsilon_\infty\varepsilon_0 \bm{m}^{\ast}m_0)$, where $q$ denotes the charge, $\varepsilon_0$ is the vacuum permittivity, and $\varepsilon_{\infty}$ denotes the high frequency dielectric constant. The cyclotron frequency tensor is defined as $\bm{\omega}_{\mbox{{\tiny c}}}=qB/(m_{0})\bm{m}^{\ast-1}$. The external magnetic field is given by $\bm{B}=B(b_{1}, b_{2}, b_{3})$ with $|\bm{B}|=B$. In general, the OHE allows independent determination of the FCC concentration parameter $N$, the mobility tensor $\bm{\mu}$, and the effective mass tensor $\bm{m}^{\ast}$, as well as the sign of the charge $q$. Without applied magnetic field, Eq.~\ref{DA} is equivalent to a diagonal tensor with the classical Drude expression on the diagonal elements.

 In order to reduce parameter correlation, the MIR-SE measurements are obtained at multiple angles of incidence and all data is analyzed simultaneously. A regression analysis\,(Levenberg-Marquardt algorithm) is performed, where the model parameters are varied until calculated and experimental data match as closely as possible (best-match model). 

\section{Results and Discussion}

\subsection{Phonon modes and infrared dielectric function}\label{Sec:phonons}

\begin{figure} 
	\centering
		\includegraphics[keepaspectratio=true,trim= 0 0 0 0, clip, width=0.48\textwidth]{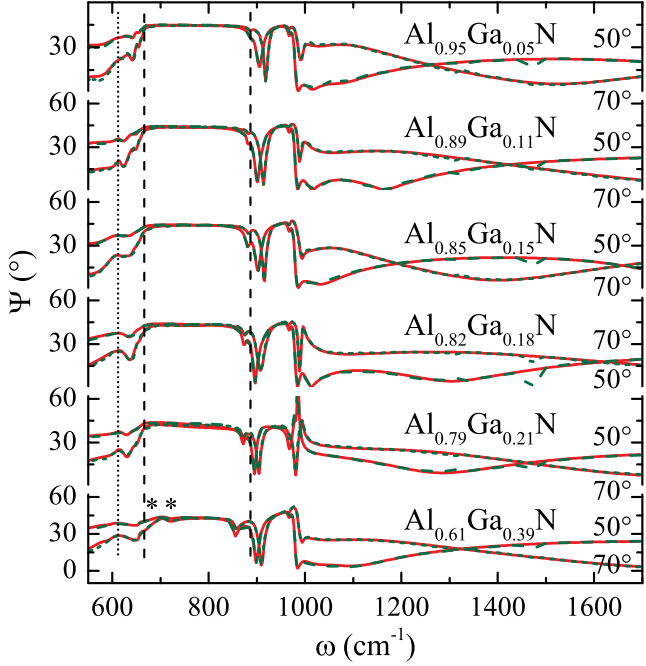} 
	\caption{Experimental (dotted lines) and best-match model calculated (solid lines) spectra of the standard ellipsometric parameter $\Psi$ for the 4H-SiC substrate/AlN buffer/graded AlGaN/undoped Al$_{x}$Ga$_{1-x}$N/Si-doped Al$_{x}$Ga$_{1-x}$N sample structures in the spectral range from 600\,cm$^{-1}$ to 1700\,cm$^{-1}$ for two angles of incidence of  $\varPhi_{\stext{a}}=50^{\circ}$ and $\varPhi_{\stext{a}}=70^{\circ}$. The Al and Ga content is indicated for each data set. The AlN $E_{1}$(TO) and $A_{1}$(LO) phonon frequency positions are marked with vertical dashed lines. The vertical dotted line is a guide to the eye and marks the approximate position of the GaN-like $E_{1}$(TO) phonon mode in the low-Al-content AlGaN layers. Two additional modes were included in the MDF in order to match the experimental data for the sample with $x=0.61$ which are marked by asterisks. } 
	\label{fig:IRSE}
\end{figure}

Figure~\ref{fig:IRSE} depicts experimental and best-matched spectra of the standard ellipsometry parameters $\Psi$ for all samples, obtained at angles of incidence of $\varPhi_{\stext{a}}=50^{\circ}$ and $\varPhi_{\stext{a}}=70^{\circ}$. The experimental (broken lines) and best-match model calculated (solid lines) data are found to be in excellent agreement. The spectra are dominated by the typical features of group-III nitride layers on SiC substrates in the reststrahlen range of the sample constituents~\cite{Kasic03,SchocheAPL13}. The frequency positions of the AlN $E_{1}$(TO) and $A_{1}$(LO) phonons are marked with vertical dashed lines in Fig.~\ref{fig:IRSE}.  
Features that can be attributed to the Al$_{x}$Ga$_{1-x}$N layers are observed in the spectra which shift continuously with increasing the Al content in the layers. The most prominent additional feature in the $\Psi$ spectrum is a dip structure on the low-frequency side of the AlN $A_1$(LO) mode which shows a strong dependence on the Al content $x$ in the Al$_{x}$Ga$_{1-x}$N layers. The AlN-like $E_1$(TO) mode only shifts slightly to lower frequencies with decreasing Al content. An additional feature appears in the spectra on the low-frequency side of the AlN-like $E_1$(TO) mode that can be assigned to the GaN-like $E_1$(TO) mode. This feature only shifts slightly with increasing Al content. A guide to the eye (dotted line) marks the approximate position of this mode in Fig.~\ref{fig:IRSE}. The feature related to this mode becomes more prominent with decreasing Al content indicating an increasing oscillation strength. Additional oscillation structures between the GaN-like and AlN-like $E_1$(TO) mode as seen for some of the samples can be attributed to thickness oscillations within the undoped and Si-doped Al$_{x}$Ga$_{1-x}$N layers. These oscillations only appear in samples of relative large AlGaN layer thicknesses (see summary in Tab.~\ref{tab:AlGaN_summary}). The determination of both $E_1$(TO) modes in these heterostructures from the MIR-SE measurements is possible since both phonon modes are located at frequencies below the reststrahlen range of the AlN buffer layer, the graded AlGaN layer and the 4H-SiC substrate.~\footnote{Precise determination of the $E_1$(TO) modes in high-Ga-content Al$_x$Ga$_{1-x}$N by MIR-SE in Ref.~\citenum{Kasic03} was impeded since these modes were energetically situated within the reststrahlen range of the GaN buffer layer and did not result in distinct spectral features.} The high peaks observed above the reststrahlen range of the AlGaN layer for the sample with Al content of $x=0.79$ in Fig.~\ref{fig:IRSE} is related to a very high FCC concentration parameter of $1.1\times10^{19}$\,cm$^{-3}$ as determined from the model analysis. 

A MDF containing two TO-LO phonon mode pairs of $E_1$ symmetry and one phonon mode pair of $A_1$ symmetry for the DF tensor components $\varepsilon_\perp$ and $\varepsilon_\parallel$, respectively, suffices to excellently match the experimental data for all samples except the one of the lowest Al content $x=0.61$. For this sample, two additional low-polarity modes had to be included at $\omega^{\mathrm{AM1}}_{\perp}=689.9$\,cm$^{-1}$ and $\omega^{\mathrm{AM2}}_{\perp}=710.6$\,cm$^{-1}$ in order to achieve a good match between experimental and model data as marked by the asterisks in Fig.~\ref{fig:IRSE}. The sensitivity for the out-of-plane high-frequency dielectric constant $\varepsilon_{\infty,\parallel}$ was insufficient to independently vary this parameter during the regression analysis. A constant anisotropy according to the results for the AlN film was assumed for all samples, i.e., a constant offset of 0.05 compared to the in-plane high-frequency dielectric constant $\varepsilon_{\infty,\perp}$ was assumed for $\varepsilon_{\infty,\parallel}$ for all samples. FCC contributions were allowed in the Si-doped top layer only and in order to determine the FCC concentration $N$ and the optical mobility parameter $\mu$, the value of the isotropically averaged effective mass parameter for each Al$_{x}$Ga$_{1-x}$N sample was varied according to the Al content $x$ (see Fig. \ref{Fig:AlGaN-effective-mass}). The sensitivity for the optical mobility parameter $\mu_\parallel$ along the $c$-axis was too low to independently determine values for $\mu_\parallel$ and $\mu_\perp$. Therefore an isotropically averaged optical mobility parameter $\mu$ was assumed for the model calculations. The determined phonon mode parameters and FCC parameters for all Al$_{x}$Ga$_{1-x}$N samples are summarized in Table I.

\begin{figure*} 
	\centering
			\includegraphics[keepaspectratio=true,trim= 0 0 0 0, clip, width=0.45\textwidth]{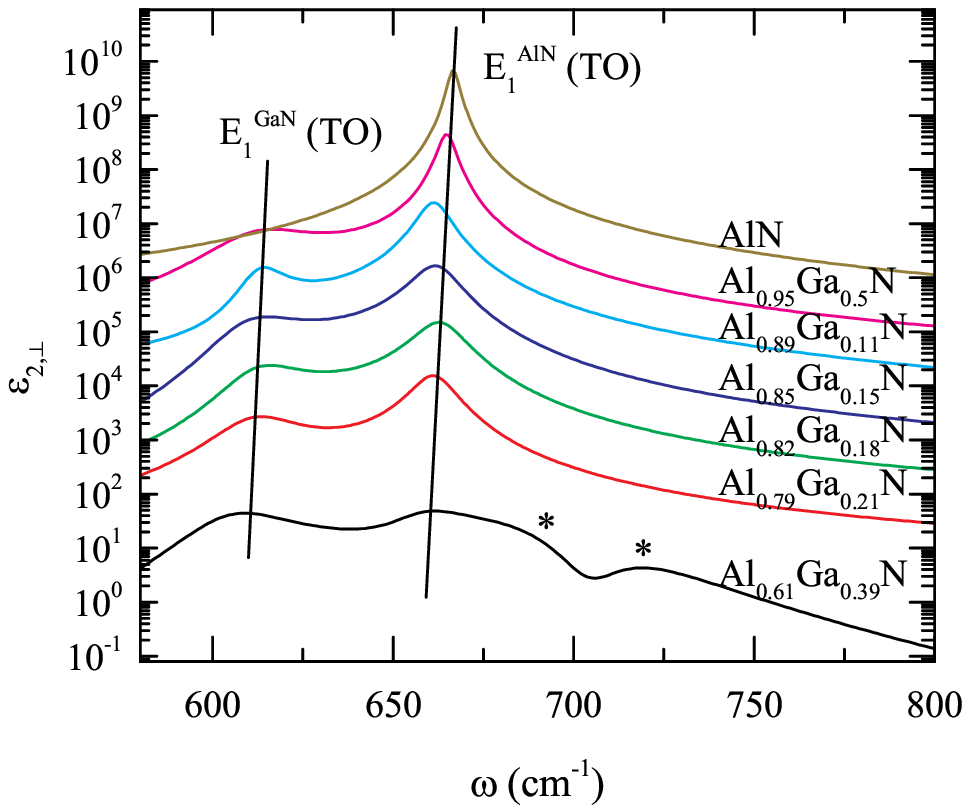} 
			\includegraphics[keepaspectratio=true,trim= 0 0 0 0, clip, width=0.45\textwidth]{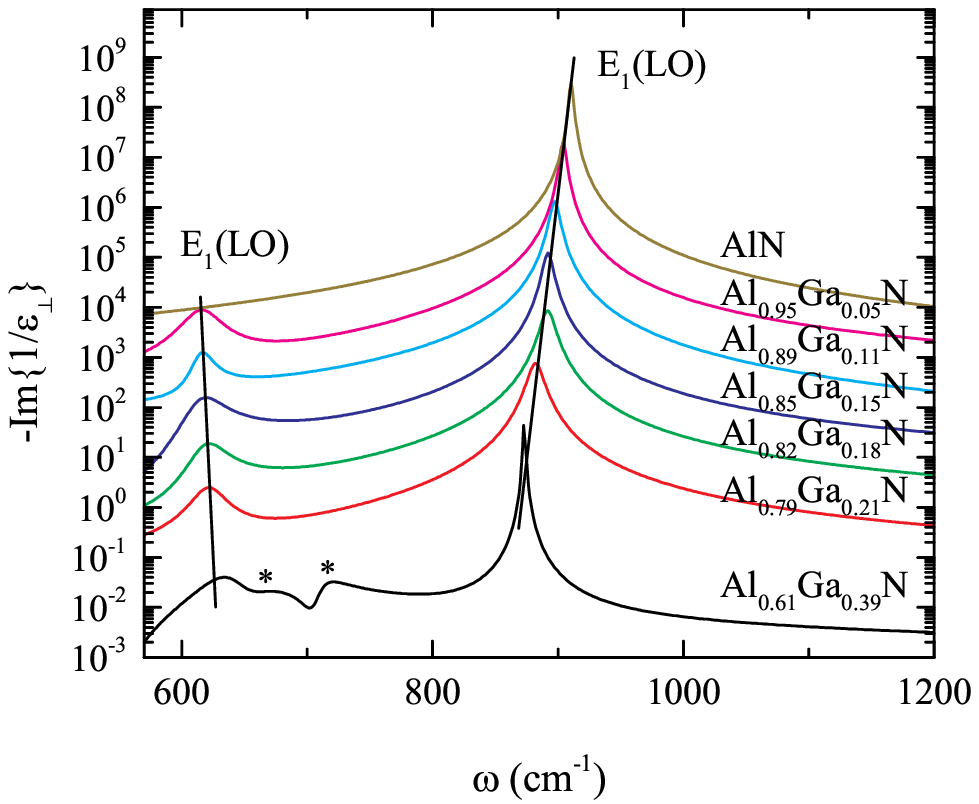} 
			\caption{Imaginary part of the dielectric function tensor component $\varepsilon_\perp$ (left) and the imaginary part of the dielectric loss function 1/$\varepsilon_\perp$ (right) for Al$_{x}$Ga$_{1-x}$N layers of different Al contents $x$ as derived from the best-match model analysis. Peak positions correspond to the E$_1$(TO) resonances (left) and E$_1$(LO) resonances (right), respectively. The determined dielectric function tensor components describe the lattice term $\varepsilon^\mathrm{L}$ without FCC contributions. The LO phonon mode frequencies are therefore equivalent to the uncoupled LO phonon modes without phonon-plasmon coupling effects. Data sets are offset by a factor of 10 for clarity. Straight solid lines are guides to the eye. Note, the different line-shape for the sample with $x\,=\,0.61$ which is a result of including two additional modes (marked by asterisks) necessary to match the experimental data. } 
	\label{fig:epsx-2}
\end{figure*}

Fig.~\ref{fig:epsx-2} shows the imaginary part of the DF tensor component $\varepsilon_\perp$ and of the dielectric loss function $1/\varepsilon_\perp$ as determined from the parameterized model analysis excluding the FCC-related Drude term.
The observed maxima in $\varepsilon_{2,\perp}$ mark the two $E_1$(TO) modes necessary to match experimental and model data (Fig.~\ref{fig:epsx-2}). It is seen that distinct blue-shifts of both $E_1$(TO) modes occur with increasing Al content $x$. An increase of the amplitude of the peak associated with the AlN-like $E_1$(TO) mode and a decrease of the amplitude of the peak associated with the GaN-like $E_1$(TO) mode is found for increasing Al content. An increasing broadening of the AlN-like $E_1$(TO) mode is indicated by an increasing width at half maximum of the peak with decreasing Al content, while no systematic change of the broadening is observed for the GaN-like phonon mode peak. The observed maxima in the imaginary part of the dielectric loss function $1/\varepsilon_\perp$ mark the spectral position of the associated LO phonon modes included in the parameterized model approach. A red shift of the low-frequency $E_1$(LO) mode and a blue-shift of the high-frequency $E_1$(LO) mode with increasing Al content is found. No systematic change of the peak broadening is observed. 

\begin{figure} 
	\centering
		\includegraphics[keepaspectratio=true,trim= 0 0 0 0, clip, width=8.5cm]{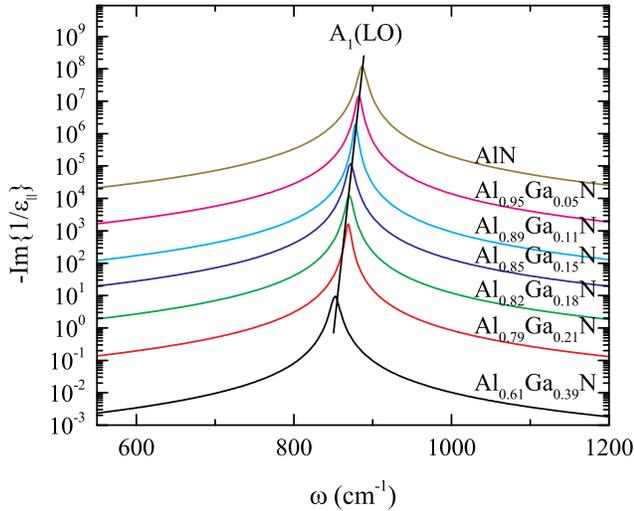}
		\caption{Imaginary part of the dielectric loss function 1/$\varepsilon_\parallel$ for Al$_{x}$Ga$_{1-x}$N layers of different Al content $x$ as derived from the best-match model analysis. Peak positions correspond to the A$_1$(LO) resonances.  The determined dielectric function tensor components describe the lattice term $\varepsilon^\mathrm{L}$ without FCC contributions. The LO phonon mode frequencies are therefore equivalent to the uncoupled LO phonon modes without phonon-plasmon coupling effects. The straight solid line is a guide to the eye. Data sets are offset by a factor of 10 for clarity.} 
	\label{fig:AlGaN-eps2z}
\end{figure}

The imaginary part of the dielectric loss function $1/\varepsilon_\parallel$ is presented in Fig.~\ref{fig:AlGaN-eps2z}. The maxima of the observed peaks mark the $A_1$(LO) phonon mode frequency. A blue-shift of the phonon mode frequency with increasing Al content $x$ is found (indicated by the guide to the eye). A systematic change in the broadening behavior is not observed for the investigated sample set. The $A_1$(TO) phonon mode is not accessible by MIR-SE for c-plane orientated group-III nitride films,~\cite{Kasic2000} and therefore a plot of the imaginary part of $\varepsilon_\parallel$ is omitted. During the model analysis and for further calculations the values of $A_1$(TO) in dependence of the Al content $x$ were adopted from Ref.~\citenum{Davydov2002}:
\begin{equation}
A_1(\mathrm{TO})=531.8+64.5(1-x)+1.9(1-x)x\;.
\label{Eq:A1-TO}
\end{equation}

\begin{figure}[tbp]
   \centering
   \includegraphics[keepaspectratio=true,trim=0 0 0 0, clip, width=0.48\textwidth]{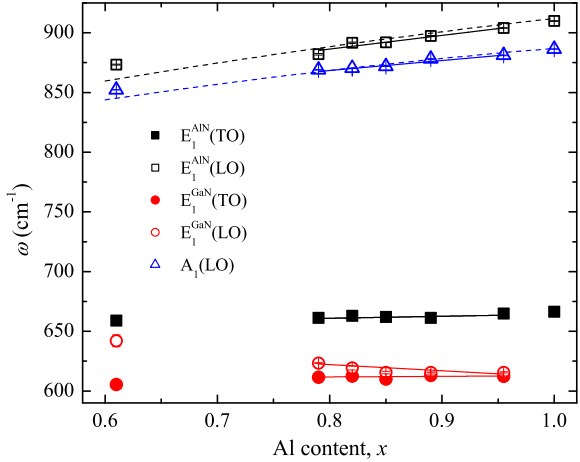} 
  \caption{Best-match model determined frequencies (symbols) of the two phonon mode pairs with $E_1$ symmetry and the single $A_1$(LO) mode in dependence of the Al content $x$. Trends for the change of the phonon mode frequency in dependence of $x$ are given as linear interpolations for high Al content by solid lines. The dependences of the AlN-like $E_1$(TO) and the $A_1$(LO) modes on the Al content $x$ as reported by Davydov~\emph{et al.} from Raman scattering measurements are given by dashed lines~\cite{Davydov2002}. }
  \label{Fig:allAlGaNphonons}
\end{figure}

The phonon mode frequencies as determined from the best-match model analysis are summarized in dependence of the Al content $x$ in Fig.~\ref{Fig:allAlGaNphonons}.  A blue-shift of the GaN-like $E_1$(TO), AlN-like $E_1$(TO), $A_1$(LO) and AlN-like $E_1$(LO) phonon mode frequencies with increasing Al content is found. Our results on the compositional dependencies of the AlN-like $E_1$(TO) and $A_1$(LO) phonon frequencies agree very well with the previously published findings from Raman spectroscopy\cite{Davydov2002} (included for comparison in Fig.~\ref{Fig:allAlGaNphonons} with dashed lines).

 In order to model the DF of the material, a TO phonon mode must always be accompanied by a LO mode (Sec.~III.B). However, a LO mode associated with the GaN-like $E_1$(TO) mode was not seen in the Raman measurements. The wave vector $\boldsymbol{q}$ of a TO phonon is oriented perpendicular to the electric field vector $\boldsymbol{E}$ and $\mathrm{div}\,\boldsymbol{E}=0$, while for propagation of LO phonons with $\boldsymbol{q}\parallel\boldsymbol{E}$ the condition Re$\left\{\varepsilon\right\}=0$ holds. In crystals with more than one phonon mode pair the second condition might not be met for all phonon mode pairs, e.g., if a weak phonon mode is energetically located close to a strong phonon mode. Such case can be interpreted as coupling of both LO modes by their macroscopic electric fields to display an effective one-mode behavior. This might explain why a two-mode behavior is reported from Raman measurements for the E$_1$(TO) modes in Al$_{x}$Ga$_{1-x}$N alloys but a one-mode behavior is found for the E$_1$(LO) modes. The splitting of associated TO and LO phonon mode pairs is a measure of the oscillation strength of the lattice vibration. 
The AlN-like $E_1$ phonon mode pair TO-LO splitting is decreasing with increasing the Al content. On the other hand, a decrease  of the TO-LO splitting for the GaN-like $E_1$  mode is determined from the model analysis. This behavior indicates a shift of spectral weight from the GaN-like phonon mode pair to the AlN-like phonon mode pair as expected for increasing Al content $x$, i.e., the oscillation strength of the GaN-like mode decreases while the oscillation strength of the AlN-like mode increases. A linear trend for high Al content $x$ for all phonon modes determined from the MIR-SE data analysis can be given as\footnote{The determination of a second order bowing parameter is not possible due to the limited composition range.}:
\begin{align}
\omega^{\mathrm{GaN}}_{TO,\perp}(x)&=611.8-4.9(1-{x})\;,\notag\\
\omega^{\mathrm{GaN}}_{LO,\perp}(x)&=611.7+5.2(1-{x})\;,\notag\\ 
\omega^{\mathrm{AlN}}_{TO,\perp}(x)&=664.3-16.7(1-{x})\;,\label{Eq:AlGaNphonons}\\
\omega^{\mathrm{AlN}}_{LO,\perp}(x)&=909.2-114.3(1-{x})\;,\notag\\
\omega_{LO,\parallel}(x)&=885.1-83.4(1-{x})\;.\notag
\end{align}

\begin{figure*}[tbp]
\centering
			\includegraphics[keepaspectratio=true,trim= 0 0 0 0, clip, width=0.45\textwidth]{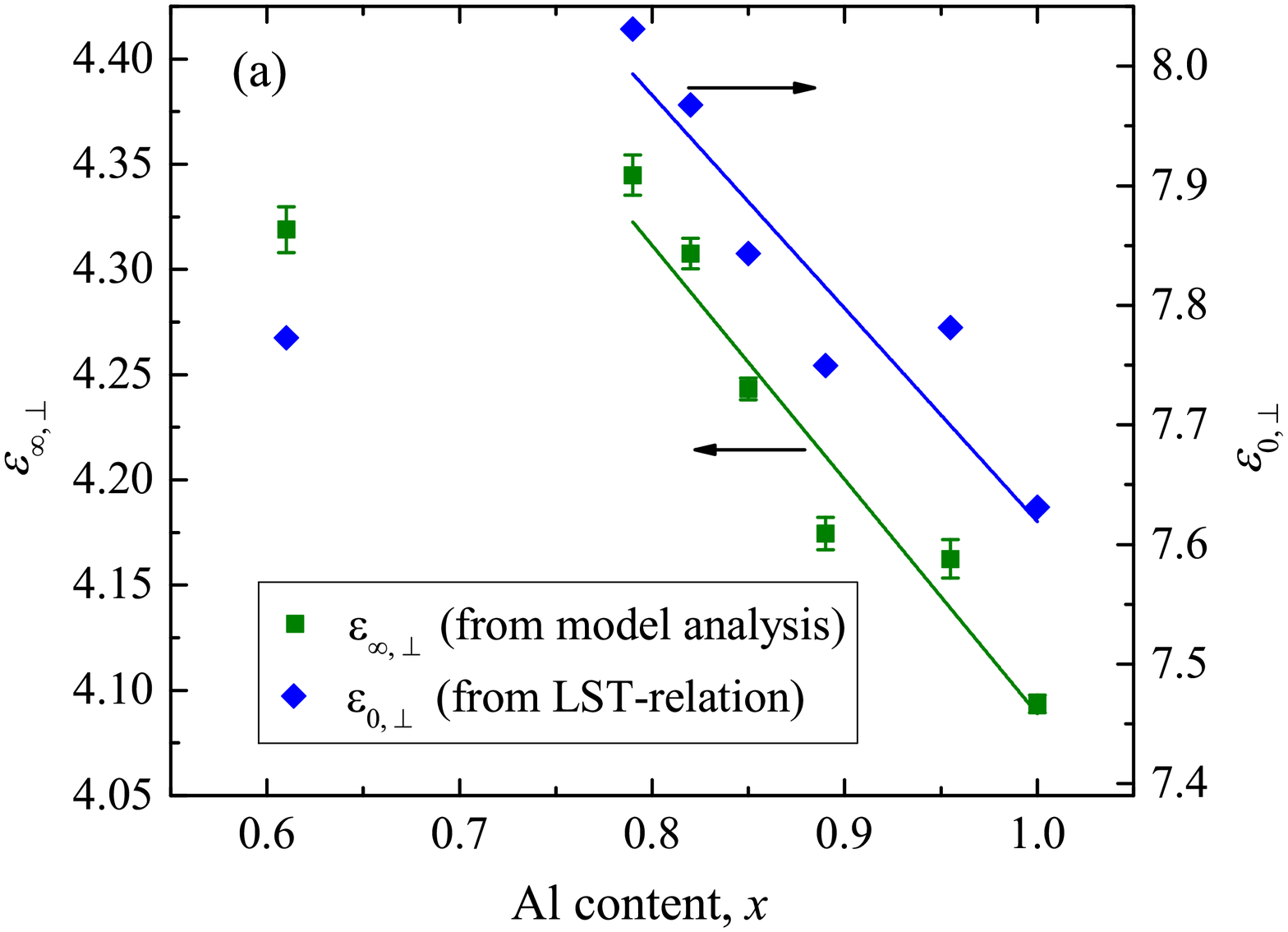} 
			\includegraphics[keepaspectratio=true,trim= 0 0 0 0, clip, width=0.45\textwidth]{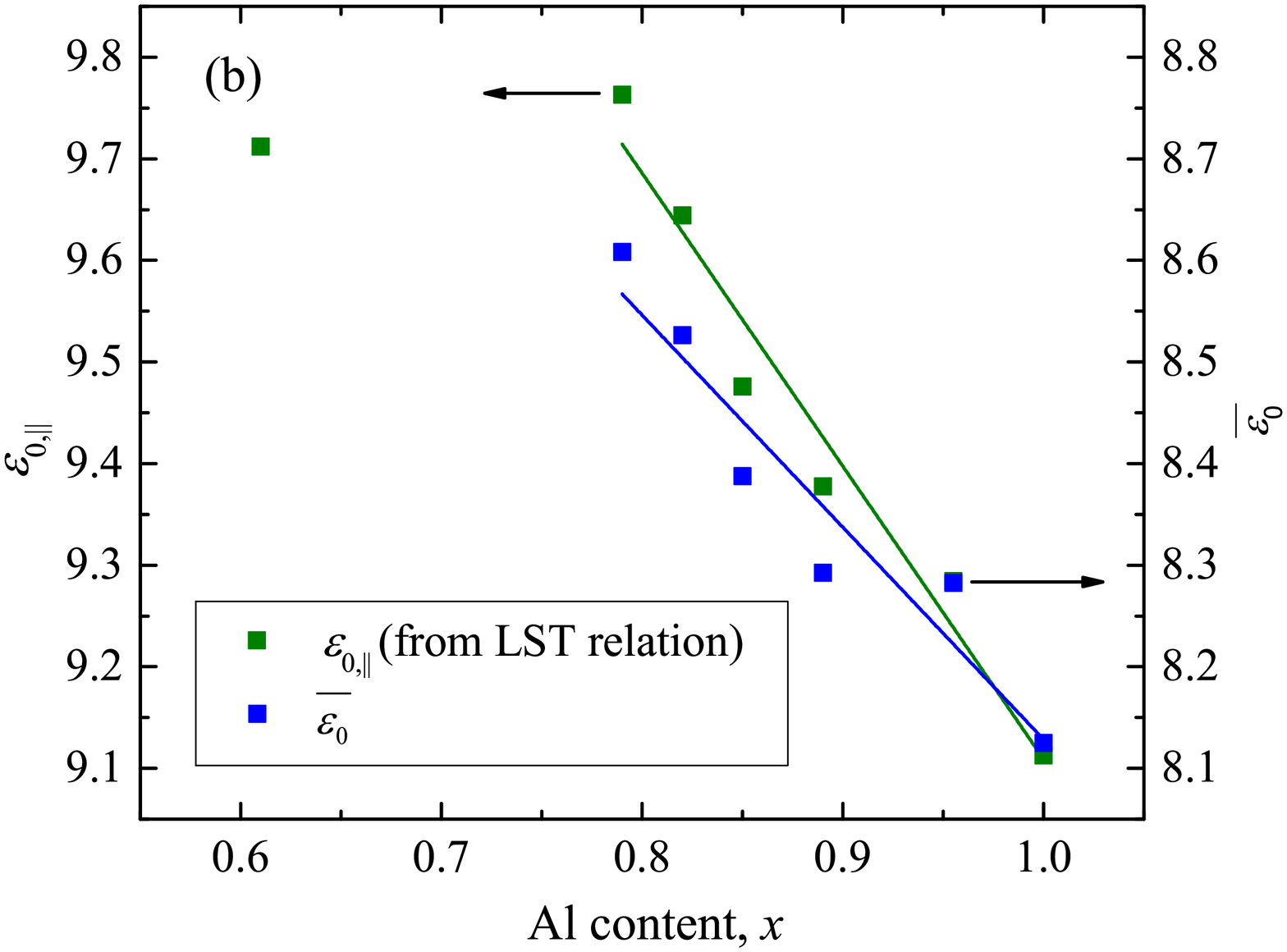}  
\caption{(a) Best-match model determined high-frequency dielectric constant $\varepsilon_{\infty,\perp}$ (left scale) and static dielectric constant $\varepsilon_{0,\perp}$ as calculated from the LST relation (right scale) in dependence of the Al content $x$. (b) Static dielectric constant $\varepsilon_{0,\parallel}$ as calculated from the LST relation (left scale) by using the $A_1$(TO) mode compositional dependence from Ref.~\citenum{Davydov2002} and the average static dielectric constant $\bar{\varepsilon_0}$ (right scale) in dependence of the Al content $x$. The respective linear dependencies of the parameters on the Al content $x$ are indicated in (a) and (b).}
  \label{Fig:AlGaN-staticeps}
\end{figure*}

The static dielectric constants $\varepsilon_{0,\perp}$ for the Al$_{x}$Ga$_{1-x}$N alloys were calculated by using the best-match model determined phonon mode parameters and high-frequency dielectric constants $\varepsilon_{\infty,\perp}$ and by applying the LST relation (Eq.\ref{Eq:LST}). The best-match model determined high-frequency dielectric constants $\varepsilon_{\infty,\perp}$ and calculated static dielectric constants $\varepsilon_{0,\perp}$ in dependence of the Al content $x$ are presented in Fig.~\ref{Fig:AlGaN-staticeps}(a). A systematic decrease with increasing the Al content is observed for both parameters. Only the sample with the lowest  Al content $x=0.61$ does not fit in the linear trend which might be related to a reduced crystalline quality. For instance, alloy disorder, compositional fluctuations or increased defect density and strain may deteriorate crystal quality as the Ga content increases.  This speculation is further supported by the observation of additional symmetry-forbidden modes in the MIR DF of this sample, that might be disorder activated (see discussion on silent $B_1$(high) mode below). The static dielectric constant $\varepsilon_{0,\parallel}$ is calculated by using the value of the high-frequency dielectric constant $\varepsilon_{\infty,\parallel}$ with an offset of 0.05 compared to $\varepsilon_{\infty,\perp}$ and by using the LST relation with the $A_1$(TO) phonon mode values calculated according to Eq.~\ref{Eq:A1-TO}. The resulting dependence of $\varepsilon_{0,\parallel}$ and the averaged static dielectric constant $\bar{\varepsilon_0}$\,=\,$\frac{1}{3}$(2$\varepsilon_{0,\perp}$+$\varepsilon_{0,\parallel}$) on the Al content $x$ is shown in Fig.~\ref{Fig:AlGaN-staticeps}(b). Similar linear dependence on $x$ as for the static dielectric constants $\varepsilon_{0,\perp}$ is found.
The dependence of the high-frequency dielectric constant $\varepsilon_{\infty,\perp}$, the static dielectric constants $\varepsilon_{0,\perp}$ and $\varepsilon_{0,\parallel}$, and the average static dielectric constant $\bar{\varepsilon_0}$ on $x$ for the range of Al contents studied are given by  the following linear equations:
\begin{align}
\varepsilon_{\infty,\perp}&=4,089+1.1(1-{x})\;, \notag\\
\varepsilon_{0,\perp}&=7.619+1.8(1-{x})\;,\label{Eq:eps-inf}\\
\varepsilon_{0,\parallel}&=9.109+2.9(1-{x})\;, \notag\\
\bar{\varepsilon_0}&=8.13+2.1(1-{x})\;.\notag
\end{align}

\begin{figure}[tbp]
\centering
   \includegraphics[keepaspectratio=true,trim=0 0 0 0, clip, width=0.48\textwidth]{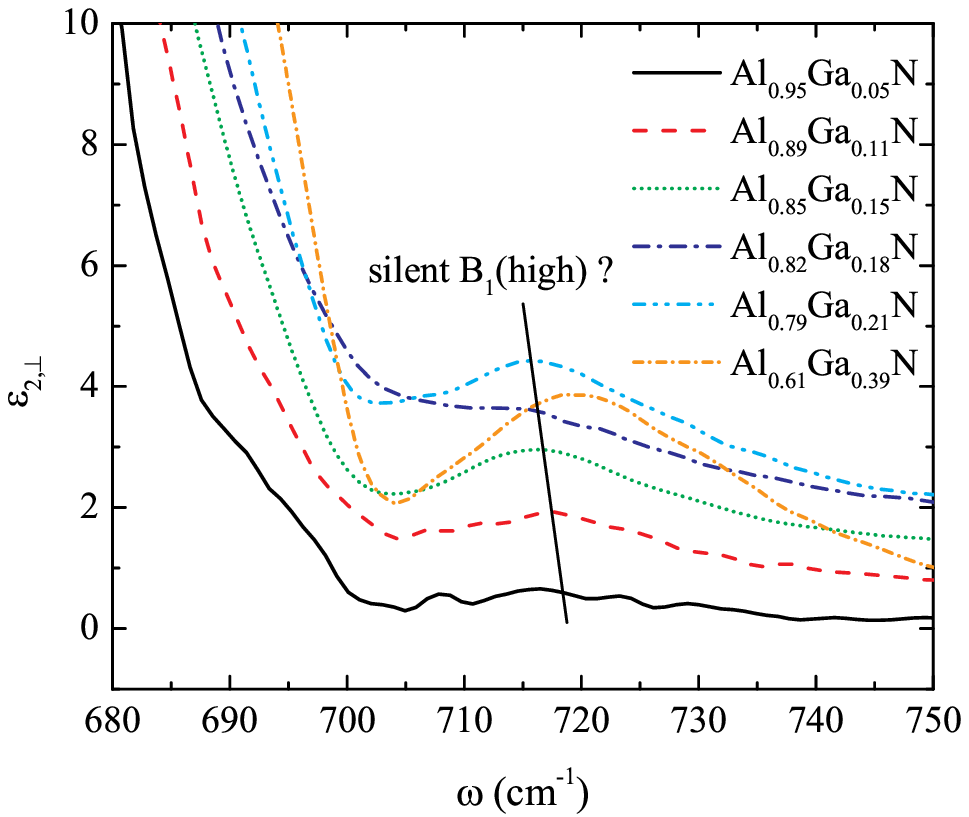} 
   \caption{Point-by-point extracted DF tensor component $\varepsilon_\perp$ around the additional mode AM2 as found in the sample with Al content $x=0.61$. A guide to the eye is included indicating a shift of the mode to lower frequencies with increasing $x$.}
  \label{Fig:B1-silent}
\end{figure}

For the sample with $x=0.61$ two additional modes had to be included at $\omega^{\mathrm{AM1}}_{\perp}=689.9$\,cm$^{-1}$ and $\omega^{\mathrm{AM2}}_{\perp}=710.6$\,cm$^{-1}$ for good match between experimental and model data. The position of these additional modes is marked by asterisks in Figs.~\ref{fig:IRSE} and~\ref{fig:epsx-2}. A more detailed inspection of the experimental data of all investigated samples in the spectral range around these additional modes reveals slight mismatch between the experimental and best-match model calculated data around the higher-frequency additional mode in all samples of this set. Except for the sample of the lowest Al content $x=0.61$, a regression analysis including the additional modes did not result in stable results. Therefore, a wavelength-by-wavelength analysis was performed in the spectral range around the higher-energy mode AM2. During this procedure, only the DF tensor component $\varepsilon_\perp$ was matched wavelength-by-wavelength while all other parameters as determined in the best-matching model were kept fixed. The resulting DF tensor component $\varepsilon_\perp$ is shown in Fig.~\ref{Fig:B1-silent}. A red-shift of the maximum with decreasing Al content $x$ is observed. A similar feature and trend was reported by Davydov~\emph{et al.} from first- and second-order Raman scattering and identified as the silent mode $B_1$(high)~\cite{Davydov2002}. Davydov~\emph{et al.} proposed that disorder in the alloy activates this otherwise silent mode. The significantly higher broadening seen for the sample with $x=0.61$ in Fig.~\ref{fig:epsx-2} and~\ref{fig:AlGaN-eps2z} and significantly different values of the high-frequency dielectric constant (Fig.~\ref{Fig:AlGaN-staticeps}) suggest that a decrease of the crystalline quality accompanied by composition fluctuations and disorder might also activate this silent mode in our sample set. 

\begin{table*}[tbp]
  {\centering
		\begin{minipage}{0.98\textwidth}
  \caption{Summary of best-match model determined parameters of all investigated samples in dependence of the molar Al content $x$. The table includes the determined concentration of Si in the doped Al$_{x}$Ga$_{1-x}$N layer as determined from SIMS, the net dopant concentration $N_\mathrm{D}-N_\mathrm{A}$ as determined from C-V measurement, the thickness parameters of the AlN buffer layer and doped Al$_{x}$Ga$_{1-x}$N layer, frequency and broadening parameters of all accessible optical phonon modes, the high-frequency dielectric constants $\varepsilon_{\infty}$, the FCC parameter concentration $N$, mobility $\mu$, and effective electron mass $m^\ast$ as determined or assumed in the model analysis.}
	\renewcommand{\arraystretch}{1.1}
	\centering
    \begin{tabular}{@{\extracolsep{7pt}}cccccccc}
    \hline
		\hline  \\ [-2.3ex]
    $x$ (XRD) & 1   &   0.955 & 0.89 & 0.85 & 0.82 & 0.79 &  0.61 \\
    \hline \\ [-2.3ex]
    $\left[\mathrm{Si}\right]$ (cm$^{-3}$) & 0      & low 10$^{17}$ & $1.1\times10^{19}$ &   low 10$^{17}$ &  $1.9\times10^{18}$ &  $9.6\times10^{18}$ &   $1.9\times10^{18}$ \\
    $N_\mathrm{D}-N_\mathrm{A}$ (cm$^{-3}$) &       & $5.4\times10^{17}$ &  $6.2\times10^{18}$ &  $4.9\times10^{17}$ &   $3.2\times10^{18}$ &   $9.6\times10^{18}$ &     $2.0\times10^{18}$ \\
      $t^{\mathrm{AlN}}$ (nm)  & $1179\pm2$  & $681\pm26$  & $630\pm6$  & 	$766\pm5$  & $509\pm5$  & $727\pm3$  & 	$395\pm5$ \\
      $t^{\mathrm{AlGaN}} $ (nm)  &     & 	$1006\pm3$  & $512\pm4$  & 	$1157\pm5$  & $370\pm3$  & $484\pm5$  & 	$838\pm2$ \\
    $\omega_{\mathrm{TO},\perp}^{\mathrm{AlN}}$ (cm$^{-1}$)      & $666.4\pm0.2$ & 	$664.9\pm0.2$ & 	$661.3\pm0.1$ & 	$662.0\pm0.2$ & $663.1\pm0.3$ & $661.3\pm0.2$ & 	$659.1\pm1.6$ \\
    $\omega_{\mathrm{LO},\perp}^{\mathrm{AlN}}$ (cm$^{-1}$)      & $909.9\pm0.3$  & 	$904.1\pm0.2$ & 	$897.4\pm0.5$ & 	$892.0\pm0.1$ & $891.2\pm0.5$ & $882.1\pm0.6$ & 	$873.3\pm0.4$ \\
    $\gamma_{\mathrm{TO},\perp}^{\mathrm{AlN}}$ (cm$^{-1}$)      & $3.1\pm0.2$ & 	$4.9\pm0.2$ & 	$9.1\pm0.2$ & 	$12.8\pm0.3$ & $13.8\pm0.5$  & $13.2\pm0.3$ & 	$22.8\pm3.2$ \\
    $\gamma_{\mathrm{LO},\perp}^{\mathrm{AlN}}$ (cm$^{-1}$)      & 3.1  & 	$4.9\pm0.2$     & 	$7.4\pm0.8$   & 	$7.9\pm0.2$ & $11.0\pm0.8$ & $13.2\pm0.3$ & 	$14.0\pm0.9$ \\
    $\omega_{\mathrm{TO},\perp}^{\mathrm{GaN}}$ (cm$^{-1}$)      &       & $612.3\pm0.6$      & 	$613.0\pm0.4$ & 	$609.9\pm0.9$ & $612.4\pm1.5$  & $611.5\pm0.6$& 	$605.4\pm1.2$ \\
    $\omega_{\mathrm{LO},\perp}^{\mathrm{GaN}}$ (cm$^{-1}$)      &       & 	$615.6\pm0.4$ & 	$615.5\pm0.5$ & 	$615.4\pm1.1$  & $619.3\pm1.6$  &$ 623.4\pm0.3$& 	$642.1\pm4.9$ \\
    $\gamma_{\mathrm{TO},\perp}^{\mathrm{GaN}}$ (cm$^{-1}$)      &       & 	$28.9\pm4.3$  & 	$11.3\pm0.6$ & 	$23.2\pm2.0$  & $21.4\pm2.8$ & $28.1\pm1.0$ & 	$30.5\pm2.1$ \\
    $\gamma_{\mathrm{LO},\perp}^{\mathrm{GaN}}$ (cm$^{-1}$)      &       & 	$33.2\pm4.4$  & 	$15.3\pm0.7$ & 	$31.1\pm2.1$  & $28.6\pm2.8$  & $36.3\pm1.0$ & 	$44.2\pm9.4$ \\
    $\omega_{\mathrm{LO},\parallel}$ (cm$^{-1}$)      & $886.4\pm0.1$ &$881.1\pm0.3$& 	$878.1\pm0.2$ & 	$871.9\pm0.1$& $870.3\pm0.1 $& $869.1\pm0.2$ & 	$852.2\pm0.1$ \\
    $\gamma_{\mathrm{LO},\parallel}$ (cm$^{-1}$)      &$ 9.1\pm0.1$ & $6.6\pm0.4$ & 	$5.2\pm0.4$ & 	$8.5\pm0.1$ & $8.2\pm0.2$ & $6.3\pm0.4$ & 	$9.8\pm0.2$\\
    $\varepsilon_{\infty,\perp}$ & $4.08\pm0.02$ & 	$4.16\pm0.02$ & 	$4.17\pm0.03$& 	$4.24\pm0.02$ & $4.30\pm0.03$ & $4.34\pm0.04$ & 	$4.31\pm0.05$\\
    $\varepsilon_{\infty,\parallel}$ & $4.13\pm0.04 $& \footnote{Insufficient sensitivity for this parameter in the regression analysis, the same anisotropy as for AlN is assumed, a constant offset of 0.05 is added to $\varepsilon_{\infty,\perp}$.}     &  \footnotemark[1]         &    \footnotemark[1]      & \footnotemark[1]         &     \footnotemark[1]      &   \footnotemark[1]     \\
    $N$ ($10^{18}$\,cm$^{-3}$)       &       & $0.28\pm0.02$ &$2.3\pm0.2$ & $0.26\pm0.02$ &$ 2.4\pm0.1$&$11\pm1$ &$ 2.9\pm0.1$ \\
    $\mu_\perp$ (cm$^2$/Vs) &       &	$7\pm2$  & 	$12\pm1$  & 	$33\pm2 $  & $32\pm1$  &$39\pm1$ & $	26\pm1 $\\
    $\mu_\parallel$ (cm$^2$/Vs) &     &\footnote{Insufficient sensitivity for this parameter in the regression analysis, isotropic behavior of this parameter is assumed.}    &  \footnotemark[2]     &\footnotemark[2]      &   \footnotemark[2]     & $32\pm1 $& \footnotemark[2]     \\
    $m^\ast_\perp$ ($m_0$) &          & 0.358 \footnote{Linearly extrapolated value from results of the OHE measurements.}&      0.350\footnotemark[3]  &       0.344\footnotemark[3]  &     $0.339\pm0.006$ & $0.334\pm0.010$ & 0.313\footnotemark[3]  \\
    $m^\ast_\parallel$ ($m_0$) &           & \footnotemark[2]  &   \footnotemark[2]&  \footnotemark[2]  & \footnotemark[2]   &$ 0.338\pm0.025$ & \footnotemark[2]\\
		\\ [-2.3ex]
    \hline
		\hline
    \end{tabular}%
	\end{minipage}
	\label{tab:AlGaN_summary}}%
\end{table*}%

\subsection{Free charge carrier parameters}\label{Sec:fcc}

\begin{figure}[tb] 
	\centering
			\includegraphics[keepaspectratio=true,trim=0 0 0 0, clip, width=0.48\textwidth]{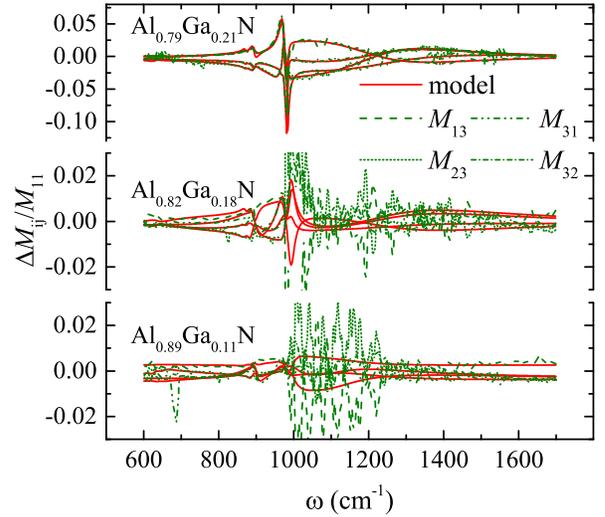}
			\caption{Experimental (broken lines) and best-model calculated (solid lines) Mueller matrix difference spectra [$\Delta M_{ij}=M_{ij}(B=+7.0~$T$)-M_{ij}(B=-7~$T)] for the samples with Al content $x$=0.89, 0.82, and 0.79 in the spectral range from 600\,cm$^{-1}$ to 1700\,cm$^{-1}$ obtained at an angle of incidence $\varPhi_{\stext{a}}=45^{\circ}$ for magnetic field orientation along the incoming light beam ($B_\mathrm{z}=(1/\sqrt{2})B$). Note the different scale for the sample of $x=0.79$.}
	\label{fig:OHE}
\end{figure}

The best-match model FCC parameters of all investigated samples in section IV.A are summarized in Table I. The table also includes the Si concentration in the doped Al$_{x}$Ga$_{1-x}$N layer and the net dopant concentration $N_\mathrm{D}-N_\mathrm{A}$. For a subset of samples with $x$\,=\,0.89, 0.82 and 0.79, MIR OHE measurements were performed in order to investigate the composition dependence of the electron effective mass parameter in high-Al-content Al$_{x}$Ga$_{1-x}$N. Note that the MIR analysis rendered no FCC contributions in the SiC and GaN indicating that the OHE signal is solely related to the FCC in the Si-doped Al$_{x}$Ga$_{1-x}$N layer. The experimental and best-match model calculated data for these samples are presented in Fig.~\ref{fig:OHE}. Difference spectra of the off-diagonal block Mueller matrix spectra are presented for measurements with magnetic field strengths of $B=\pm7$\,T along the incoming light beam. The off-diagonal block Mueller-matrix elements $M_{13}$, $M_{23}$, $M_{31}$ and $M_{32}$ depict the magnet-field induced birefringence and render the OHE. The lattice term of the DF, $\varepsilon^{\text{L}}_j(\omega)$, is symmetric under magnetic field inversion while the FCC term $\varepsilon^{\text{FC}}(\omega)$ contains antisymmetric frequency-dependent contributions. Therefore, the non-zero values seen in the differences of $M_{13}$, $M_{23}$, $M_{31}$ and $M_{32}$ in Fig.~\ref{fig:OHE} are solely caused by the FCC related magneto-optical birefringence~\cite{Hofmann2008a}. This magneto-optical birefringence is strongest close to the LO-phonon plasmon coupled (LPP) modes which is seen as the prominent features in Fig.~\ref{fig:OHE}. The spectral position of these features is mainly determined by the FCC concentration while the shape is influenced by the  mobility and effective mass parameters. Much stronger FCC related birefringence is found for the sample with $x= 0.79$ compared to the other two samples as seen by the much higher maximum values of the off-diagonal block Mueller matrix differences in Fig.~\ref{fig:OHE}. This indicates a higher FCC concentration and mobility parameter in this sample compared to those in the Al$_{x}$Ga$_{1-x}$N films with $x$\,=\,0.89 and 0.82. The noise above the reststrahlen range can be attributed to the loss in reflectivity in this spectral range due to DF values close to 1 resulting in low intensity on the detector for these wavelengths. The FCC related birefringence vanishes outside the range of the reststrahlen bands. Note, that the five frequency-independent parameters FCC concentration $N$, effective masses $m^\ast_\parallel$ and $m^\ast_\perp$, and mobilities $\mu_\parallel$ and $\mu_\perp$ are sufficient to achieve the excellent match between experimental (broken line) and model data (solid lines) over the entire spectral range shown in Fig.~\ref{fig:OHE}. Further note, that the data for the sample with Al content $x=0.79$ was published in Ref.~\citenum{SchocheAPL13}, but labeled with $x=0.72$ as determined from SIMS\footnotemark[1]. 

The best-match model data in Fig.~\ref{fig:OHE} was calculated from the same model used for the MIR-SE analysis during the simultaneous analysis of the MIR-SE and MIR-OHE data sets. The best-match model extracted FCC parameters were included in Table I.  All Al$_{x}$Ga$_{1-x}$N films were found from the signature in the OHE data ($\Delta M_{ij}$ becomes -$\Delta M_{ij}$ if p-type conductive) to be $n$-type conductive as expected. The five frequency-independent parameters FCC concentration $N$, effective masses $m^\ast_\parallel$ and $m^\ast_\perp$, and mobilities $\mu_\parallel$ and $\mu_\perp$ were included during the regression analysis of the sample with molar Al content $x= 0.79$. A high free electron concentration of $N=(1.1\pm0.2)\times10^{19}$~cm$^{-3}$ was determined. This value is in excellent agreement with the net dopant concentration as determined by the C-V measurements (see Table I).  No significant anisotropy of the effective mass parameter is found for this FCC concentration. The determined effective mass parameters are $m^\ast_\perp=(0.334\pm0.010)\,m_0$ perpendicular to the $c$-axis and $m^\ast_\parallel=(0.338\pm0.025)\,m_0$ parallel to the $c$-axis. The sensitivity of the analysis for the effective mass parameter for polarization parallel to the $c$-axis $m^\ast_\parallel$ is slightly reduced compared to the sensitivity for $m^\ast_\perp$. However, from the model analysis, it can be estimated that the anisotropy of the effective electron mass parameter may not be higher than 0.03. Theoretical investigations of the effective mass parameter for AlN do not predict a significant anisotropy of the effective mass at the $\Gamma$-point~\cite{Rinke2008}. The experimental result for this sample of relatively high-Al-content confirm the theoretical predictions and is also in agreement with reports of vanishing effective mass parameter anisotropy for similar electron concentrations in InN~\cite{Hofmann2008} and GaN~\cite{Kasic2000}. Slightly different mobility parameters of $\mu_\perp=(39\pm1)$\,cm$^2$/Vs and $\mu_\parallel=(32\pm1)$\,cm$^2$/Vs are determined for the directions perpendicular and parallel to the $c$-axis, respectively. Anisotropic mobility parameters have been previously observed in GaN~\cite{Kasic2000} and InN~\cite{Hofmann2008} and could be related to different distributions of extended defects or impurities in directions parallel and perpendicular to the $c$-axis. 

\begin{figure}[tbp]
   \includegraphics[keepaspectratio=true,trim=0 0 0 0, clip, width=0.48\textwidth]{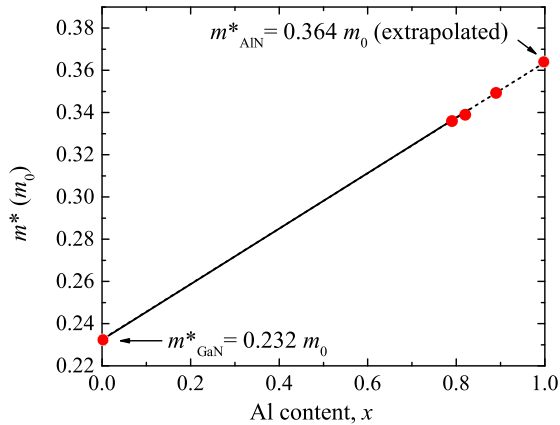} 
   \caption{Linear extrapolation of the electron effective mass parameter in Al$_{x}$Ga$_{1-x}$N alloys as estimated for assumed values of the isotropically averaged effective mass parameter of $m^\ast=0.232\,m_0$ for electrons in GaN~\cite{Kasic2000} and the OHE determined isotropically averaged value for Al$_{0.79}$Ga$_{0.21}$N of $m^\ast=0.336\,m_0$. The values as determined or assumed in the best-match model calculated OHE data in Fig.~\ref{fig:OHE} are marked by symbols.}
  \label{Fig:AlGaN-effective-mass}
\end{figure}

Under the assumption of a linear increase of the effective mass parameter in AlGaN alloys with increasing Al content and assuming an isotropically averaged effective mass parameter value of $m^\ast=0.232\,m_0$ for electrons in GaN~\cite{Kasic2000}, an isotropically averaged electron effective mass parameter value of $m^\ast=0.364\,m_0$ can be extrapolated for AlN (Fig.~\ref{Fig:AlGaN-effective-mass}). The value for the electron effective mass parameter for arbitrary values of the Al content can be calculated using the following linear trend:
\begin{equation}
m^\ast=0.364-0.131(1-{x})\;.
\label{Eq:effective-mass}
\end{equation}
This extrapolated value for the electron effective mass parameter in AlN fits well into the range between 0.28\,$m_0$ to 0.45\,$m_0$ as estimated from cathodoluminescence by Silveira~\emph{et al.}~\cite{Silveira2004}. The value is slightly larger than theoretically predicted value.\cite{Rinke2008} However, it is known that \emph{ab initio} density-functional theory calculations used in Ref.~\onlinecite{Rinke2008} tend to give underestimated values of electronic parameters. 

Due to the much smaller Mueller matrix elements and resulting smaller signal-to-noise ratio, the sensitivity during the regression analysis for the FCC parameter of the DF tensor component $\varepsilon_\parallel$ was significantly decreased for the sample with Al content $x=\,0.82$ compared to the sample with $x=\,0.79$. Therefore, isotropically averaged effective mass and mobility parameters had to be assumed for this sample in the analysis. The results for the sample with $x=\,0.79$ suggest this averaging procedure is feasible since no significant anisotropy is expected. The reduced FCC-induced birefringence is related to the much smaller free electron concentration parameter of $N=(2.4\pm0.1)\times10^{18}$~cm$^{-3}$ found for this sample. The determined isotropically averaged electron effective mass parameter $m^\ast=(0.339\pm0.006)$\,$m_0$ matches excellently with the value determined from the linear extrapolation in Fig.~\ref{Fig:AlGaN-effective-mass}. The regression analysis for the sample with Al content $x=\,0.89$ did not result in stable FCC parameters even when assuming isotropically averaged values for the effective mass and mobility parameters. The best-match model in Fig.~\ref{fig:OHE} was calculated by assuming an isotropically averaged electron effective mass parameter $m^\ast=0.350\,m_0$ as extrapolated from the results of the sample with $x=\,0.79$. The resulting excellent match between best-match model calculated and experimental data indicates that the assumed value for the effective mass parameter must be close to the actual material value. A free electron concentration of $N=(2.3\pm0.2)\times10^{18}$\,cm$^{-3}$ and mobility parameter of $\mu=(12\pm1)\,$cm$^2$/Vs are determined from the simultaneous analysis of the MIR-SE and MIR-OHE data. The significantly lower mobility parameter in the Al$_{0.89}$Ga$_{0.11}$N film (see Table I) explains the small off-diagonal block Mueller matrix elements in the OHE data of this sample.

\begin{figure}[tbp]
   \includegraphics[keepaspectratio=true,trim=0 0 0 0, clip, width=0.48\textwidth]{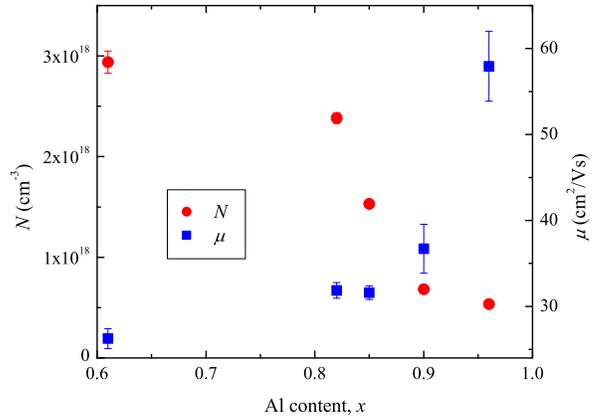} 
   \caption{Best-match model determined parameters concentration $N$ and isotropically averaged mobility $\mu$ in Si-doped Al$_{x}$Ga$_{1-x}$N alloys of similar Si concentration in the low $10^{18}$\,cm$^{-3}$ range in dependence of the Al content $x$. Effective mass parameter values according to the extrapolation in Eq.~\ref{Eq:effective-mass} were assumed during the model analysis of MIR-SE data.}
  \label{Fig:AlGaN-FCC}
\end{figure}

In order to investigate the influence of the Al content $x$ on the free electron parameters, a set of samples with similar Si concentrations of approx. $2\times10^{18}$\,cm$^{-3}$ and grown under the same conditions was investigated by MIR-SE. Effective mass parameter values according to the extrapolation in Eq.~\ref{Eq:effective-mass} were assumed during the model analysis. The resulting best-match model calculated FCC parameters are given in Fig.~\ref{Fig:AlGaN-FCC}.
A decrease of the free electron concentration with increasing Al content $x$ is found from the model analysis. These results corroborate a recent report on increasing resistivity in Al$_{x}$Ga$_{1-x}$N:Si with increasing Al content.~\cite{Nilsson2015} In principle, such behavior would be expected since the binding energy (ionization energy) of impurities in Al$_{x}$Ga$_{1-x}$N alloys increases with increasing $x$. More importantly, stable Si-related DX centers with increasing activation energy of the silicon donor are found above $x$\,=\,0.84.\cite{Trinh2014,Nilsson2015} An increase of the mobility parameter with increasing Al content $x$ is observed in Fig.~\ref{Fig:AlGaN-FCC}. In principle, ionized impurity scattering is expected to be the limiting scattering mechanism in highly doped semiconductors. The scattering probability is proportional to the number of ionized impurities. As indicated by the decreasing free electron concentration with increasing $x$, the number of ionized impurities in the investigated samples decreases with increasing $x$. Therefore, an increase in the mobility parameter with increasing Al content $x$ can be expected. However, additional factors, such as, an increased defect density and structural deterioration that may occur with increasing Ga content, may also play a role.

\section{Conclusions}

The phonon mode parameters and anisotropic MIR DF tensor components of high-Al-content Al$_{x}$Ga$_{1-x}$N alloys in dependence of the Al content $x$ were determined from MIR-SE measurements for a set of high-quality Si-doped Al$_{x}$Ga$_{1-x}$N films on 4H-SiC substrates. A subset of samples with high free electron concentration was identified and MIR-OHE measurements were performed in order to determine the electron effective mass parameter in high-Al-content Al$_{x}$Ga$_{1-x}$N alloys. Further, the composition dependencies of the free electron concentration and mobility parameters of films with similar Si-dopant concentrations were investigated and discussed.

Two-mode behavior of the $E_1$(TO) modes and one-mode behavior of the $A_1$(LO) mode is found from the  MIR-SE measurements. The composition dependencies of the phonon modes frequencies with increasing Al content $x$ are established and found to be in good agreement with previous Raman scattering spectroscopy reports.  A decreasing TO-LO splitting for the GaN-like $E_1$ mode and increasing TO-LO splitting of the AlN-like $E_1$ mode are observed for increasing $x$. Equations for the linear approximation of the phonon mode frequencies for high-Al-content $x$ are given. The static dielectric constants $\varepsilon_{0,\perp}$, $\varepsilon_{0,\parallel}$, and average static dielectric constant in dependence of $x$ are obtained by using the best-match model derived phonon mode frequencies and the high-frequency dielectric constant $\varepsilon_{\infty,\perp}$ and applying the Lydanne-Sachs-Teller (LST) relation. A systematic decrease with Al content $x$ is found and equations describing the linear composition dependence for high $x$ are given. The best-match model determined values for the uncoupled phonon mode frequency, broadening parameters and high-frequency dielectric constants allow to reproduce the MIR dielectric function tensor of high-Al-content AlGaN alloys in dependence of the Al content and will be useful for analysis of MIR-SE/reflectometry spectra of doped AlGaN alloys and device heterostructures. The reported static dielectric function and effective mass values are required for theoretical calculations in order to accurately predict and optimize device performance. An additional mode is identified in a wavelength-by-wavelength analysis of the DF tensor component $\varepsilon_{\perp}$ and attributed to a disorder-activated, silent $B_1$(high) mode. 

The electron effective mass parameters in a Al$_{0.79}$Ga$_{0.21}$N are determined from the OHE measurements to be $m^\ast_\perp=(0.334\pm0.010)\,m_0$ perpendicular to the $c$-axis and $m^\ast_\parallel=(0.338\pm0.025)\,m_0$ parallel to the $c$-axis. No significant effective mass parameter anisotropy is found in agreement with theoretical predictions. By assuming an average effective mass parameter of $m^\ast=0.232\,m_0$ for electrons in GaN and linear dependence of the electron effective mass parameter on the Al content $x$, a linear equation for the effective mass parameter in Al$_{x}$Ga$_{1-x}$N alloys as a function of $x$ is derived: $m^\ast=0.364-0.131(1-{x}$). This dependence is confirmed by OHE measurements on Al$_x$Ga$_{1-x}$N with different Al content $x$. A systematic decrease of the FCC concentration with increasing Al content $x$ is found for samples of comparable Si-dopant levels in agreement with Ref.\onlinecite{Nilsson2015}. The observed decrease is explained by the increasing binding energy of the donors with increasing $x$ and the formation of stable Si-related DX centers with increasing activation energy of the silicon donor for $x$\,=\,0.84.\cite{Trinh2014} An increase in the mobility parameter is also observed with increasing Al-content as a result of decreasing number of ionized impurities. Our results clearly demonstrate sufficient n-type doping and mobility parameters in high-Al content Al$_{x}$Ga$_{1-x}$N alloys that can open the way of their application in a myriad of electronic and optoelectronic devices.

\section{Acknowledgements}

This work is supported by the Swedish Research Council (VR) under Grant No. 2013-5580, the Swedish Governmental Agency for Innovation Systems (VINNOVA) under the VINNMER international qualification program, Grant No. 2011-03486, the Swedish Government Strategic Research Area Grant in Materials Science "Advanced Functional Materials", and the Swedish Foundation for Strategic Research (SSF), under Grants No. FFL12-0181 and RIF14-055. A.~K.~G. acknowledges support through projects VR 621-2013-5818, VINNOVA 2011-01329, and Linnaeus Initiative for Novel Functional Materials (LiLi-NFM, VR). The authors further acknowledge financial support from the National Science Foundation under awards MRSEC DMR-0820521, MRI DMR-0922937, DMR-0907475, and EPS-1004094.

%

\end{document}